\documentclass{desyproc}
\usepackage{subfig,multirow,array}
\usepackage[table]{xcolor}
\usepackage{cite}
\newcolumntype{C}[1]{>{\centering\let\newline
\\\arraybackslash\hspace{0pt}}m{#1}}
\newcommand{\at}{\makeatletter @\makeatother}

\begin{document}
\title{Quantum regime of laser-matter interactions at extreme intensities}
\author{{\slshape Alexander Fedotov
}\\[1ex]
National Research Nuclear University "MEPhI" (Moscow Engineering Physics Institute), \\ 115409 Kashirskoe sh. 31, Moscow, Russia}
\contribID{02}\confID{099}\desyproc{DESY-PROC-2016-04}
\acronym{SFHQ 2016}
\doi  
\maketitle

\begin{abstract}
A survey of physical parameters and of a ladder of various regimes of laser-matter interactions at extreme intensities is given. Special emphases is made on three selected topics: (i) qualitative derivation of the scalings for probability rates of the basic processes; (ii) self-sustained cascades (which may dominate at the intensity levels attainable with next generation laser facilities); and (iii) possibility of breaking down the Intense Field QED approach for ultrarelativistic electrons and high-energy photons at certain intensity level.
\end{abstract}

\section{Introduction}

One of the notable trends of the last decades was unprecedented growth of laser power and intensity accessible for experimental research. As of now, the most striking progress was achieved in the (near-)optical range, which corresponds to a typical wavelength $\lambda\simeq 1\mu\text{m}$, frequency $\nu=c/\lambda\simeq 10^{15}\text{Hz}$ and photon energy $\hbar\omega\simeq 1\text{eV}$. For the state-of-the-art facilities, pulse energy is typically $W_\text{L}\simeq 0.1\text{kJ}$ and pulse duration $\tau_\text{L}\simeq 100\text{fs}$ \footnote{$1\text{fs}\equiv 10^{-15}\text{s}$ -- such small durations became attainable after invention of Chirped Pulse Amplification \cite{strickland1985compression}.}. 
As a consequence, the \emph{peak} laser power is huge,
\[P_\text{L}=\frac{W_\text{L}}{\tau_\text{L}}\simeq \frac{10^2\text{J}}{10^{-13}\text{s}}=10^{15}\text{W}\equiv 1\text{PW},\]
comparable to the net power produced by a large country. The \emph{average} power, though, of course remains low due to a small repetition rate $\nu_\text{R}\simeq10^{-4}\div 10\;\text{Hz}$. Assuming that the pulses are focused to diffraction limit, the expected peak intensity is estimated by (cf.
\cite{yanovsky2008ultra}):
\[ I_\text{L}=\frac{P_\text{L}}{R^2}\simeq \frac{P_\text{L}}{\lambda^2}\simeq\frac{10^{15}\text{W}}{(10^{-4}\text{cm})^2}\simeq 10^{23}\text{W/cm}^2.\]

Furthermore, the ongoing construction or upgrade at such facilities as CLF (UK), Apollon (France), PEARL (Russia), ELI Beamlines (Czech Republic), ELI-NP (Romania), EP-OPAL (USA), QiangGuang (China) and alike is very promising in attaining laser intensities of the order of $10^{24}$W/cm$^2$ or even higher in the nearest future. Moreover, far-reaching exawatt-class facilities such as ELI \cite{ELI-site} and XCELS \cite{XCELS-site} aiming at achieving the intensity $\gtrsim 10^{26}$ W/cm$^2$ are being also planned already. All this brings reasonable prospects on further advance of experimental studies of a variety of yet unexplored phenomena of laser-matter interactions at such extreme intensities. 

\section{Basic parameters and physical regimes}

Let us discuss the key parameters of the theory, and the characteristic intensity levels corresponding to the various non-perturbative regimes of high-intensity laser-matter interaction to be compared to the experimentally accessed or expected values from the Introduction. 

Perhaps the main element of the strong field approach is a concept of external (classical) background field $\mathcal{A}_\mu$. Obviously, this implies that the number of photons in relevant field modes is not changed essentially by absorption or scattering events, in particular is huge enough. Assuming $\omega$ is characteristic frequency and estimating laser pulse energy as $W_\text{L}\simeq \frac{E^2+H^2}{8\pi}V\simeq \frac{E^2}{4\pi}V$, we arrive at $\bar{N}_\gamma\simeq W_\text{L}/\hbar\omega\gg 1$, or $E\gg \sqrt{\hbar\omega/V}\gg 1$. This criterion is always satisfied in cases of either $\omega=0$ (static field) or $V=\infty$ (infinitely extended field), but both are formally unphysical. For a tightly focused laser pulse we may assume $V\simeq \lambda^3$ which results in $E\gg \omega^2\sqrt{\hbar/c^2}$. Then the limitation on corresponding intensity reads $I_L=\frac{c}{4\pi}E^2\gtrsim 10^5\text{W/cm}^2$ and is almost ever satisfied with huge margins (it is perhaps enough to mention that typically $\bar{N}_\gamma\simeq 10^{30}\text{cm}^{-3}$), hence we adopt it wherever possible.

Another important and instructive example is a strong field concept in atomic physics. The atomic length (Bohr radius) and energy (Rydberg) are $l_\text{at}=\hbar^2/Zme^2=5.3\times 10^{-9}\text{cm}$ and $\mathcal{E}_\text{at}\simeq Ze^2/l_\text{at}= mZ^2e^4/\hbar^4\simeq 10\text{eV}$, respectively (where the particular numbers are given for hydrogen with $Z=1$). In this context the field can be considered `strong' if the work it produces across the size of an atom $eE l_\text{at}\gtrsim \mathcal{E}_\text{at}$, resulting in $E\gtrsim E_\text{at}\equiv Ze/l_\text{at}^2=m^2Z^3e^5/\hbar^4=5\times 10^9\text{V/cm}$ and $I_\text{L}\gtrsim \frac{c}{4\pi}E_\text{at}^2=3\times 10^{16}\text{W/cm}^2$. For such laser intensities material targets are rapidly ionized, thus turning into plasma. Laser-plasma interactions are usually simulated with Particle-In-Cell (PIC) codes. 

One of the most important parameters arises in a criterion of whether free electron motion driven by laser field is relativistic or not. Consider the classical equation of electron motion
\begin{equation}\label{Lorentz}
\frac{d\vec{p}}{dt}=e\left(\vec{E}+\frac{\vec{v}}{c}\times\vec{H}\right).
\end{equation}
From Eq.~(\ref{Lorentz}) the momentum of laser driven electron quiver oscillation  can be estimated as $p_\perp\simeq eE/\omega$, hence the motion is relativistic if  $a_0\equiv p_\perp/mc\simeq eE/m\omega c\gtrsim 1$ (as in this case $\gamma\simeq a_0\gtrsim 1$). This corresponds to $E\gtrsim E_\text{rel}\equiv m\omega c/e$, or $I_\text{L}\gtrsim \frac{c}{4\pi}E_\text{rel}^2\simeq 3\times 10^{18}\text{W/cm}^2$. The resulting dimensionless parameter $a_0$ admits a Lorentz- and gauge-invariant\footnote{This is indeed invariant under gauge transformations $\delta A^\mu\propto k^\mu$ of a plane wave.} definition for a plane wave field, $a_0=\frac{e}{mc}\sqrt{-\mathcal{A}_\mu \mathcal{A}^\mu}$, and a number of alternative interpretations. For example, one point is that for $a_0\gtrsim 1$ the equation of motion (\ref{Lorentz}) also becomes essentially non-linear, thus leading e.g. to harmonic generation. Hence $a_0$ is often called the (dimensionless) classical nonlinearity parameter. Due to its importance, it is also often used in laser physics community to quantify the field strength and intensity (the latter via $a_0\approx 6\times 10^{-10}\lambda\,[\mu\text{m}]\sqrt{I_\text{L}[\text{W/cm}^2]}$). In these terms the currently attained intensity level corresponds to $a_0\simeq 10^2$.

\begin{figure}
\subfloat[]{\includegraphics[width=0.6\textwidth]{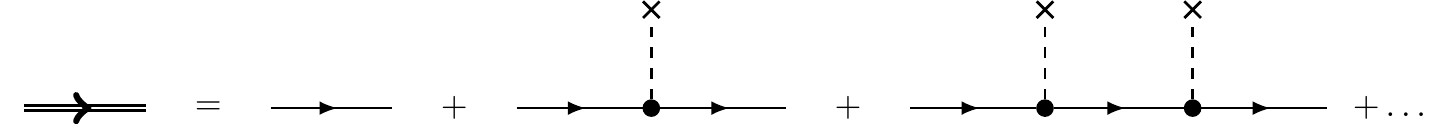}}
\hspace{1cm}
\subfloat[]{\includegraphics[width=0.3\textwidth]{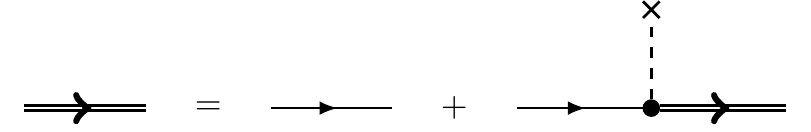}}
\caption{Exact propagator in external field (a) and a closed-form equation it obeys (b).}\label{fig:exact_prop}
\end{figure}

Yet another interpretation arises in the framework of QED, where motion of an electron in a given external field $\mathcal{A}_\mu$ is described by a sum of diagrams shown in Fig.~\ref{fig:exact_prop}a. Here each vertex corresponds to the factor $-ie\gamma^\mu\mathcal{A}_\mu$, while each thin electron line (free propagator) to $iS_0=\frac1{\gamma p-m}\simeq 1/m$. Hence, the expansion parameter of the QED perturbation theory is given by $e\mathcal{A}/m$, i.e. by the same parameter $a_0$ as above. In this context, $a_0\gtrsim 1$ corresponds to non-perturbative (also often called multiphoton) with respect to $\mathcal{A}_\mu$ interaction regime. This can be understood in a pictorial way by noticing that in presence of a large number of background photons in the relevant field mode the interaction vertex weight ($\sim\sqrt{\alpha}$ in vacuum) should acquire an additional Bose factor $\sqrt{\bar{N}_\gamma(-1)}\approx \sqrt{\bar{N}_\gamma}$,
\begin{equation}\label{vertex}
\sqrt{\alpha}\to \sqrt{\alpha}\times\sqrt{\bar{N}_\gamma}\simeq \sqrt{\alpha}\times \sqrt{l_C^2\times\lambda\times \bar{n}_\gamma}\simeq\frac{e}{\sqrt{\hbar c}}\times\sqrt{\left(\frac{\hbar}{mc}\right)^2\times\frac{2\pi c}{\omega}\times\frac{E^2}{4\pi\hbar\omega}}\simeq a_0,
\end{equation}
and hence is effectively replaced by $a_0$. Here we assumed that the electron probes the photons contained in an effective tubular interaction volume of Compton width $\simeq l_C=\hbar/mc$ around the classical electron trajectory of length $\lambda$. Note that $\hbar$ totally cancels, thus assuring that $a_0$, even though being an expansion parameter of QED perturbation theory, is nevertheless still of purely classical origin.

Since the focus of the rest of the paper is on laser-matter interactions at the high-power laser facilities, we always assume below that $a_0\gg 1$. Under such condition all-orders summation with respect to interaction with external field should be done (see Fig.~\ref{fig:exact_prop}a). This is naturally done within the approach which I call the Intense Field QED (IFQED). Namely, it is easy to see that the `exact' (with respect to interaction with external field), or `dressed', electron propagator obeys the equation shown schematically in Fig.~\ref{fig:exact_prop}b. The rules for computation of the QED amplitudes are then formulated as in ordinary QED, but with free fermion lines and propagators replaced with the dressed ones. This approach was tested experimentally many times indirectly in atomic physics as well as directly in the late 90's famous E144 SLAC experiment \cite{Burke1997,Bamber1999}.

When laser intensity is further increasing, novel physical regimes should show up. For example, under the condition $a_{0i}=(Ze)E/M\omega c\gtrsim 1$, or $a_0=eE/m\omega c\gtrsim M/Zm\simeq 2M_p/m\sim 4\times 10^3$, the ions should also become relativistic. The corresponding intensity is $I_\text{L}\gtrsim 5\times 10^{25}\text{W/cm}^2$ and is far beyond the currently attainable level. Another complication should show up at even lower intensities. Assuming $\gamma\simeq a_0$ and $E_\perp,\,H\simeq E$, the radiation reaction force
\begin{equation}\label{RR}
\vec{F}_\text{rad}\simeq -\frac{2e^4 \left(\vec{E}+\frac{\vec{v}}{c}\times\vec{H}\right)_\perp^2\gamma^2}{3m^2c^5}\vec{v},
\end{equation}
acting on electron, becomes $\gtrsim eE$ for $E\gtrsim \left(m^4\omega^2c^6/e^5\right)^{1/3}$, or $a_0\gtrsim \left(mc^3/e^2\omega\right)^{1/3}\simeq 400$, which corresponds to $I_\text{L}\gtrsim 5\times 10^{23}\text{W/cm}^2$. In this regime one should account for classical radiation reaction in simulations of laser-matter interactions.  Accounting for both effects requires just adequate correction of the developed PIC codes, and this is now indeed a hot topic in the laser physics community.

Self-action of an electron was at the focus of classical theory for decades and, as well known, was one of original motivations for invention of quantum theory. Besides radiation reaction force, it also implies existence of electromagnetic contribution into electron mass, classically of the order of $\mathcal{E}_\text{em}(r_0)\simeq e^2/r_0$, where $r_0$ is the electron radius. In the limit of a pointlike electron ($r_0\to 0$) this contribution diverges, but paradoxes appear in classical theory already when the mass correction $\mathcal{E}_\text{em}(r_0)\gtrsim mc^2$, i.e. for $r_0\lesssim r_e\equiv e^2/mc^2$ ($r_e$ is called the `classical electron radius'). The radiation reaction force (\ref{RR}) may produce across the distance $r_e$ the work $\gtrsim mc^2$ if the field strength in a proper reference frame exceeds $E_\text{P}\gtrsim E_\text{cr}\equiv m^2c^4/e^3$. These conditions are considered as limits of applicability of Classical Electrodynamics. 

\setlength\intextsep{0pt}
\begin{wraptable}{l}{0.48\textwidth}
\subfloat[]{\begin{tabular}{|c|p{0.18\textwidth}|p{0.12\textwidth}|}\hline
Regime	&$a_0\ll 1$&$a_0\gtrsim 1$\\\hline
$\chi\ll 1$& classical non-relativistic& classical relativistic\\\hline
\rowcolor{gray!25}
$\chi\gtrsim 1$& perturbative QED& IFQED\\\hline
\end{tabular}}\\
\subfloat[]{\begin{tabular}{|c|p{0.3\textwidth}|}\hline
$I_\text{L}[\text{W/cm}^2]$ & PHYSICAL REGIME\\\hline
\rowcolor{gray!25}$5\times 10^{29}$(?)& Sauter-Schwinger QED critical field\\
\rowcolor{gray!25}$5\times 10^{25}$& Relativistic ions\\
\rowcolor{gray!25}$2.5\times 10^{25}$& Massive self-sustained QED cascades\\
\rowcolor{gray!25}$10^{24}$& Quantum radiation reaction, pair photoproduction ($\chi\gtrsim 1$)\\
$5\times 10^{23}$& Classical radiation reaction\\ 
\hline$5\times 10^{22}$& \emph{Mid'2010s state-of-the-art}\\\hline
$3\times 10^{18}$& Relativistic electrons ($a_0\gtrsim 1$)\\
$3\times 10^{16}$& Strong field of atomic physics (rapid ionization and plasma formation)\\
$10^5$& External (given classical background) field concept\\
$<10^5$ & Week field quantum regime\\
\hline\end{tabular}}
\caption{Classification of physical regimes of laser-matter interactions according to the values of the key parameters $a_0$ and $\chi$ (a) and the ladder of various regimes successively attainable upon increasing laser intensity (b). Domain corresponding to quantum regime at extreme intensity is colored in gray.}
\label{Regimes}
\end{wraptable}

However, things are quite different in quantum theory (QED), which introduces a novel Compton scale $l_{\text{C}}=\hbar/mc\approx 137r_e$. Even though electron still formally remains pointlike (as required by relativity), it turns out it cannot be localized at size smaller that $l_{\text{C}}$ because of uncontrollable disturbance of vacuum (pair creation). For example, the average of the charge density operator in one-electron state is delocalized to a size $\simeq l_{\text{C}}$. As a consequence, the divergency of electron self energy is partially canceled by contribution of vacuum virtual pairs and becomes much weaker, $\mathcal{E}_\text{em}(r_0)\simeq (e^2/l_{\text{C}})\log\left(l_{\text{C}}/r_0\right)$. In this sense, a pointlike charge is effectively replaced by a cloud of virtual pairs of size $\simeq l_{\text{C}}=1/m\simeq 4\times 10^{-11}\text{cm}$. That is why we adopted the Compton length as the effective width of electron `trajectory' in our above estimation (\ref{vertex}). 
Unlike in Classical Electrodynamics, radiation reaction in QED can never outreach the Lorentz force (thus the known paradoxes of the former are avoided), however the work produced by the field across the distance $l_{\text{C}}$ can still exceed the rest energy, $eE_\text{S}l_{\text{C}}\simeq mc^2$. The required field strength $E_\text{S}\equiv m^2c^3/e\hbar=1.3\times 10^{16}\text{V/cm}$ (note that $E_\text{S}\simeq \frac1{137}E_\text{cr}$) is called the Sauter-Schwinger, or QED critical field, and corresponds to laser intensity $I_\text{L}=\frac{c}{4\pi}E_\text{S}^2\simeq 5\times 10^{29}\text{W/cm}^2$, which is far beyond the capabilities of the existing or prospective laser facilities. Besides laser physics, the electric and magnetic\footnote{Since magnetic fields do not produce work, for them the criticality condition is formulated e.g. by demanding the principle Landau level to be relativistic.} fields $\gtrsim E_\text{S}$ may arise in heavy ion collisions, magnetic fields $H\simeq m^2c^3/e\hbar\simeq 4\times 10^{13}\text{G}$ are also anticipated around compact astrophysical objects (magnetars). 

Since the electromagnetic field strengths are not Lorentz invariant, it may seem not obvious for which reference frame the condition $E,\,H\gtrsim E_\text{S}$ should be formulated. In fact this criterion should be formulated in terms of Lorentz invariants. For example, in absence of particles (i.e. in vacuum) the only field invariants are $E^2-H^2=-\frac12F_{\mu\nu}F^{\mu\nu}$ and $\vec{E}\cdot\vec{H}=\frac18\epsilon_{\mu\nu\lambda\varkappa}F^{\mu\nu}F^{\lambda\varkappa}$. In such a case the field strengths should actually exceed $E_\text{S}$ in a reference frame, for which either electric or magnetic fields vanishes or they are parallel. However, in presence of a particle with $4$-momentum $p^\mu$ one extra Lorentz invariant, usually called the dynamical quantum parameter, can be defined:
\begin{equation}\label{chi}
\chi=\frac{e\hbar}{m^3c^4}\sqrt{-(F_{\mu\nu}p^\nu)^2}=\frac{\gamma\sqrt{\left(\vec{E}+\frac{\vec{v}\times\vec{H}}{c}\right)^2-\frac{(\vec{v}\cdot\vec{E})^2}{c^2}}}{E_\text{S}}.
\end{equation}
It actually acquires several equivalent native physical meanings, e.g. in case of electron a ratio of the electric field strength to $E_\text{S}$ in its proper (rest) frame, or proper acceleration in Compton units. Typically, in the lab frame $E_{\parallel}\sim E_{\perp}$, where the components are denoted according to direction of particle momentum. Hence for ultrarelativistic particle $E_{\text{P}\parallel}\sim E_{\parallel}$, $E_{\text{P}\perp}\sim\gamma E_{\perp}\gg E_{\text{P}\parallel}$ and $\chi= E_\text{P}/E_\text{S}\simeq \gamma E_{\perp}/E_\text{S}$. Yet another interpretation stems from the fact that for radiating electron with $\chi\lesssim 1$ it is also the average emitted photon energy-to-electron energy ratio, hence the condition $\chi\gtrsim 1$ also indicates the significance of quantum recoil. The quantum regime of laser-matter interaction is naturally discriminated by\footnote{It should be stressed that the quantum regime of interaction with \emph{strong external classical background} under discussion arises due to recoil in essentially multiphoton radiation processes (e.g. hard photon emission), as opposed to the (completely different!) \emph{week field quantum regime} on bottom of Table~\ref{Regimes}b, where quantum effects arise due to absorption or emission of individual laser photons.} $\chi\gtrsim 1$.

Generally speaking, the introduced parameters $a_0=eE/m\omega c$ (classical nonlinearity parameter) and $\chi\simeq \gamma E_\perp/E_\text{S}$ (dynamical quantum parameter) are independent and allow for classification of various regimes of laser-matter interactions, see Table~\ref{Regimes}a. For instance, in SLAC experiment both $a_0,\,\chi\sim 1$. However, if the electrons are not accelerated by external sources but driven by the field then, assuming $E_\perp\sim E$ and $\gamma\sim a_0\gg 1$, we arrive at $\chi\simeq (\hbar\omega/mc^2)a_0^2\gtrsim 1$ for $a_0\gtrsim \sqrt{mc^2/\hbar\omega}\simeq 700$. The required laser intensity in such case would be $I_\text{L}\gtrsim 10^{24}\text{W/cm}^2$. The list of various laser-matter interaction regimes discussed above, along with the required laser intensities, given relative to the present day state-of-the-art level, is summarized in Table~\ref{Regimes}b. Further details can be found in the reviews \cite{Bulanov2006,diPiazza2012,Narozhny2015}.

\section{Qualitative analysis of basic IFQED processes}\label{sec:qual}

Within the framework of IFQED approach, calculation of probabilities of a process is reduced to calculation of diagrams with `exact', or `dressed' electron external lines and propagators (see Fig.~\ref{fig:exact_prop}a), determined by equation in Fig.~\ref{fig:exact_prop}b. The latter equation can be solved exactly for just a few particular external field backgrounds (e.g. constant field, plane wave, Coulomb field), but even when it admits exact solution, application of the method usually results in extremely bulky intermediate calculations \cite{Ritus1985}. But as a rule, qualitative considerations may result in a deeper insight into the problem. Here I am going to demonstrate how at least some of the known asymptotic expressions for probability rates of basic QED processes in strong external field could receive a simple-man derivation, based exclusively on kinematical and dimensional arguments together with the uncertainty principle (see \cite{Fedotov2015} for a more detailed presentation). 

\begin{figure}
\centering
\subfloat[]{\includegraphics[width=0.2\textwidth]{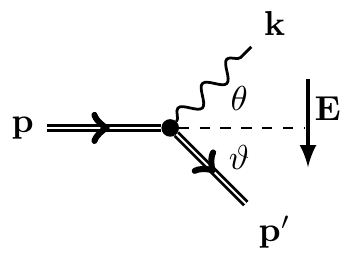}}\hspace{3cm}
\subfloat[]{\includegraphics[width=0.2\textwidth]{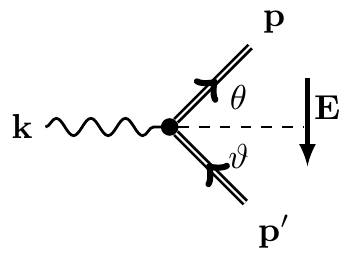}}
\caption{Basic IFQED processes: photon emission (a) and pair photoproduction (b).}
\label{fig:basic_proc}
\end{figure}

In presence of external field the QED processes can be subdivided into field-modified (i.e. those which occur even in absence of the field) and field-induced. Let us focus below on the field-induced processes only. The diagrams for the basic processes of this kind, single photon emission and pair photoproduction, are shown in Fig.~\ref{fig:basic_proc}. For such sort of processes we can introduce the energy lack $\Delta\varepsilon=\sum \varepsilon_f-\sum\varepsilon_i>0$ (where the indices refer to final and initial particles) and a couple of characteristic time scales, the first one ($t_e$) is the time it takes for the field to provide the amount of work required for a process to occur (symbolically, $e\int_0^{t_e}\vec{E}\cdot d\vec{s}\simeq\Delta\varepsilon$), and the second one ($t_q\simeq 1/\Delta\varepsilon$) is the time for which according to the uncertainty principle the final state can endure as a virtual one. As we will shortly observe, it turns out that in a relativistically strong laser field of our interest ($a_0\gg 1$) both these timescales are much smaller than the laser period $\sim 1/\omega$, hence the laser field can be considered as \emph{locally constant}. In addition, we also assume whenever possible that the particles are ultrarelativistic and moving transversely with respect to the field, in such a case the field can be also considered \emph{crossed} ($\vec{E}\perp\vec{H}$ and $E=H$), and the total probabilities depend exclusively on the quantum dynamical parameter(s) $\chi_i$. Furthermore, since \emph{any} field is all the same equivalent to a constant crossed field, we can replace it e.g. with the constant purely electric field directed orthogonally to the momentum of the initial particle. Then, by picking up the time gauge $\vec{A}(t)=-\vec{E}t$, the energies of the charged particles can be written as $\varepsilon_{\vec{p}}(t)=[(\vec{p}-e\vec{A}(t))^2+m^2]^{1/2}$.

Obviously, if $t_e\gg t_q$ then the process is (quasi-)classical, in particular its probability should be exponentially suppressed. The probability of the process under such conditions can be estimated by the quasiclassical `imaginary time' technique,
\begin{equation}\label{QC}
W_{i\to f}\propto \left\vert\exp\left\{-\int_0^{t_*} \Delta\varepsilon(it')\,dt'\right\}\right\vert^2,
\end{equation}
where $t_*$ is determined by the conditions of energy-momentum conservation $\Delta\varepsilon(it_*)=0$, $\Delta\vec{p}=0$. It turns out that typically $t_*\simeq t_e$, so that $W_{i\to f}=\mathcal{O}( e^{-t_e/t_q})$. If, on the contrary, $t_e\ll t_q$, then the process is essentially quantum and unsuppressed.

Let us first demonstrate how the scheme works using a popular example of electron-positron pair creation from vacuum. For this case $\Delta\varepsilon=2m$, $t_e\simeq m/eE$ and $t_q\simeq 1/m$. Note that for $a_0\gg 1$ we have $t_e=1/\omega a_0\ll 1/\omega$, thus the locally constant field approximation should work. For $E\ll E_\text{S}=m^2/e$ we have $t_e\gg t_q$, so that the process should be suppressed. The expected suppression factor is $e^{-t_e/t_q}\simeq e^{-E_\text{S}/E}$. More precisely, assuming\footnote{Here $\vec{p}_\perp$ denotes the component of the momentum which is orthogonal to the field.} (for the sake of simplicity only) $\vec{p}_\perp=0$ we obtain: $\Delta\varepsilon(t)=2\sqrt{m^2+e^2E^2t^2}$, $\Delta\varepsilon(it_*)=0$, $t_*=m/eE=t_e$ (as announced above), and
\begin{equation}\label{Schwinger}
W_{e^-e^+}=\left\vert\exp\left\{-2\int_0^{m/eE}\sqrt{m^2-e^2E^2{t'}^2}\,dt'\right\}\right\vert^2=e^{-\pi m^2/eE},
\end{equation}
in perfect agreement with the exact result of Sauter, Heisenberg-Euler, Schwinger and others.

As a second illustration, consider pair photoproduction by a hard photon (see Fig.~\ref{fig:basic_proc}b). For this case, the energy lack can be written as
\begin{equation}\label{dE}
\begin{aligned}
\Delta\varepsilon(t)&=\varepsilon_{\vec{p}'}(t)+\varepsilon_{\vec{p}}(t)-k\approx\sqrt{(k-p)^2+e^2E^2t^2+m^2}+\sqrt{p^2+e^2E^2t^2+m^2}-k\approx\\
&\approx k-p+\frac{e^2E^2t^2+m^2}{2(k-p)}+p+\frac{e^2E^2t^2+m^2}{2p}-k=\frac{k\left(e^2E^2t^2+m^2\right)}{2p(k-p)}\gtrsim \frac{2\left(e^2E^2t^2+m^2\right)}{k},
\end{aligned}
\end{equation}
where the minimum is attained for $p=p'=k/2$. In the weak field limit the second term in the numerator ($m^2$) should dominate over the first term ($e^2E^2t^2$) so that $\Delta\varepsilon\simeq m^2/k$. It is well known that for ultrarelativistic kinematics $\theta,\vartheta\simeq m/k\ll 1$. Hence the characteristic scales
\begin{equation*}
t_q\simeq 1/\Delta\varepsilon\simeq k/m^2\ll t_e\simeq \Delta\varepsilon/eE\vartheta\simeq m/eE,
\end{equation*}
if the dynamical quantum parameter of the initial hard photon $\varkappa=\frac{eEk}{m^3}\lesssim 1$. According to the above in such a regime we expect that $W_{e^-e^+}=\mathcal{O}(e^{-1/\varkappa})$. Indeed, looking for a stationary point we obtain $\Delta\varepsilon(it_*)=2\left(-e^2E^2t_*^2+m^2\right)/k=0$, $t_*=m/eE=t_e$, and
\begin{equation}\label{PPcl}
W_{e^-e^+}=\left\vert\exp\left(-\int_0^{m/eE}\frac{2\left(-e^2E^2t^2+m^2\right)}{k}\,dt\right)\right\vert^2=\left\vert e^{-4m^3/3eEk}\right\vert^2=e^{-8/3\varkappa}.
\end{equation}
This is again in accordance with exact computations, but [unlike Eq.~(\ref{Schwinger})] we are unaware of other simple-man derivations of this result in  previous literature.

In the opposite (quantum, $\varkappa\gg 1$) strong field limit, according to Eq.~(\ref{dE}), $\Delta\varepsilon(t)\simeq e^2E^2t^2/k$ and angular spread arises mostly due to contortion of electron and positron trajectories by the field, $\theta(t),\vartheta(t)\simeq eEt/k\ll 1$. Hence $eE\vartheta(t) t\simeq\Delta\varepsilon(t)$ identically, i.e. $t_e$ is not fixed but arbitrary. At the same time, from its definition $t_q\simeq 1/\Delta\varepsilon(t_q)\simeq k/e^2E^2t_q^2$ we obtain 
\begin{equation}\label{t_q}
t_q\simeq\left(\frac{k}{e^2E^2}\right)^{1/3}\equiv\frac{m}{eE}\varkappa^{1/3}\equiv\frac{k}{m^2\varkappa^{2/3}}.
\end{equation}
Hence, for $\varkappa\gg 1$ having just a single time scale parameter $t_q$, on dimensional grounds we have
\begin{equation}\label{W_q}
W_{e^-e^+}(\varkappa\gg 1)\simeq \frac{e^2}{t_q}\sim \frac{e^2m^2}{k}\varkappa^{2/3},\quad W_\gamma(\chi\gg 1) \simeq \frac{e^2m^2}{p}\chi^{2/3},
\end{equation}
where the second formula gives the probability $W_\gamma$ of hard photon emission by electron (see Fig.~\ref{fig:basic_proc}a) for $\chi\gg 1$, as ultrarelativistic kinematics for both processes is exactly the same as soon as electron mass is neglected. These results are (up to numerical factor $\sim 1$) also in perfect agreement with exact calculations. 

\section{Self-sustained QED cascades}

\begin{table}[t]
\centering
\begin{tabular}{|l|C{0.25\textwidth}|C{0.4\textwidth}|}\hline
\multirow{2}{*}{Property} &\multicolumn{2}{c|}{QED cascades in external field} \\\cline{2-3}
&S(hower)-type&A(valanche)-type\\\hline
Energy source & energy $\varepsilon_0$ of seed particle
&external field (donates energy by particles acceleration)\\\hline
Multiplicity $N_{e^-e^+}$ & limited by $\varepsilon_0$ & exponentially increases ($\propto e^{\Gamma t}$), potentially up to macroscopic value\\\hline
Proceeds until\ldots& \ldots secondary particles lose energy and $\chi$ & 
\parbox{0.4\textwidth}{
$\bullet$ \ldots particles escape (?);\\
$\bullet$ \ldots field depletion (?);\\
$\bullet$ \ldots thermolization (?);
}\\\hline
Similar to: & Extensive Air Showers & gas or dielectric discharge\\
\hline
\end{tabular}
\caption{S(hower)-type \emph{vs.} A(valanche)-type QED cascades in external field.}\label{SAtypes}
\end{table}

In case seed particle in a strong field has $\chi\gtrsim 1$, it can emit photons with $\varkappa\sim 1$, which are in turn capable for pair photoproduction. In such a case these events may follow one by one, thus forming a chain called a QED cascade. In principle, this is very similar to the familiar cascades generated by high energy particles in matter due to Bremsstrahlung and pair photoproduction on nuclei (e.g., as in Extensive Air Showers). However, a novel distinctive feature for cascades in a strong laser field is that such a field may not serve only a target, but in general is also capable for acceleration of charged particles, thus donating them energy and making the cascade self-sustained. The principle differences between these two types of cascades (which we abbreviate S- and A-type \cite{Mironov2014}) are summarized in Table~\ref{SAtypes}. Here our point is that production of A-type cascades starting from a certain intensity level may dominate in laser-matter interactions resulting in creation of macroscopic amount of pairs and hard photons, as shown in Table~\ref{Regimes}b, thus totally changing the landscape of laser-matter interactions \cite{Narozhny2014}. Moreover, since this process (at least in most probable scenarios) leads to depletion of external field, it can also prevent practical attainability of the Sauter-Schwinger critical field with optical lasers \cite{Fedotov2010} (this point is indicated in Table~\ref{Regimes}b by a question mark).

During the process of hard photon emission or pair photoproduction the values of energy $\varepsilon$ and dynamical quantum parameter $\chi$ of a parental particle are partitioned among the secondary ones. As stressed above, the main distinctive feature of a self-sustained (A-type) cascade against the ordinary S-type cascade is the `acceleration stage' where these $\varepsilon$ and even more importantly $\chi$ are then restored back by the field before the next QED event takes place. In contrast to consideration of a QED process itself, at this stage the difference between the actual field and a constant crossed one should be necessarily taken into account (that is why we call our constant and crossed field approximation \emph{local}), as otherwise $\chi$ would be conserved exactly. At the same time, motion of ultrarelativistic particles in between the QED events may be still considered classically\footnote{It turns out that motion in a constant crossed field is classical exactly. In a general setting, for a subcritical ($E\ll E_{\text{S}}$) electric field quasiclassical approximation breaks down only near the turning points, where particles are slow. Ultrarelativistic motion in a subcritical magnetic field is also classical since particle occupies high Landau levels.}. Concerning a problem of $\varepsilon$ and $\chi$ evolution along a particle trajectory in a generic field configuration, a non-trivial result is that (apart from a few artificial particular cases, e.g. linearly polarized purely electric field or a running plane wave field) for an initially slow particle at time scales $m/eE\ll t\ll 1/\omega$ we always have
\begin{equation}\label{acc_mech}
\varepsilon(t)=m\gamma(t)\simeq eEt,\quad E_\perp(t)\simeq E\omega t,\quad \chi(t)\simeq \frac{E_\perp(t)}{E_{\text{S}}}\gamma(t)\simeq \frac{mc^2}{\hbar\omega}\left(\frac{E}{E_{\text{S}}}\right)^2(\omega t)^2.
\end{equation} 
Hence it takes for such a particle $t_{\text{acc}}\simeq \alpha E_{\text{S}}/\omega E$ to gain $\chi\sim 1$, where for illustration purpose we have skipped the accidental dimensionless factor $\sqrt{\alpha^2mc^2/\hbar\omega}\simeq 4.5$ (square root of Rydberg in eV's).

\begin{figure}
\centering
\subfloat[]{\includegraphics[width=0.32\textwidth]{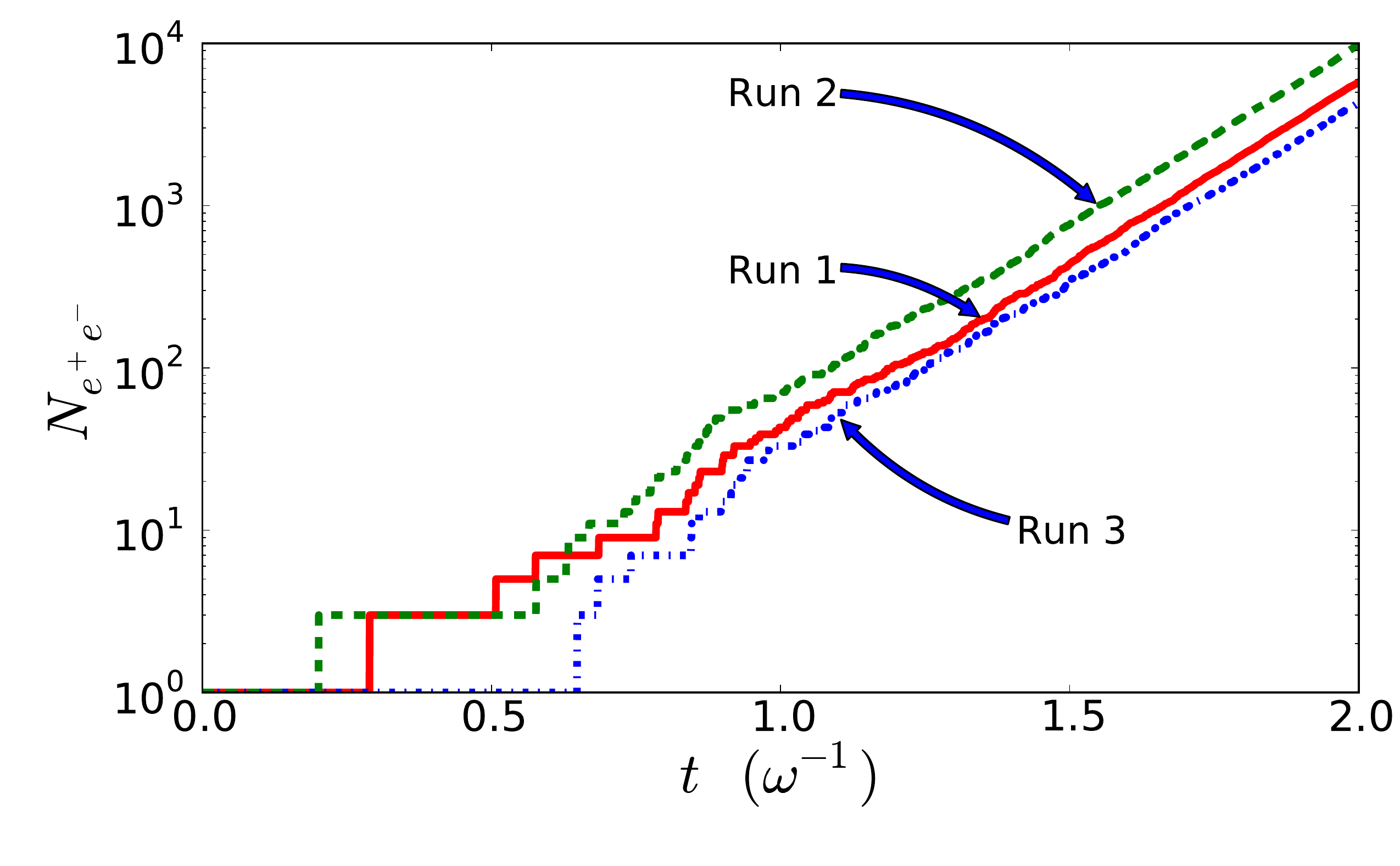}}\hspace{0.1cm}
\subfloat[]{\includegraphics[width=0.32\textwidth]{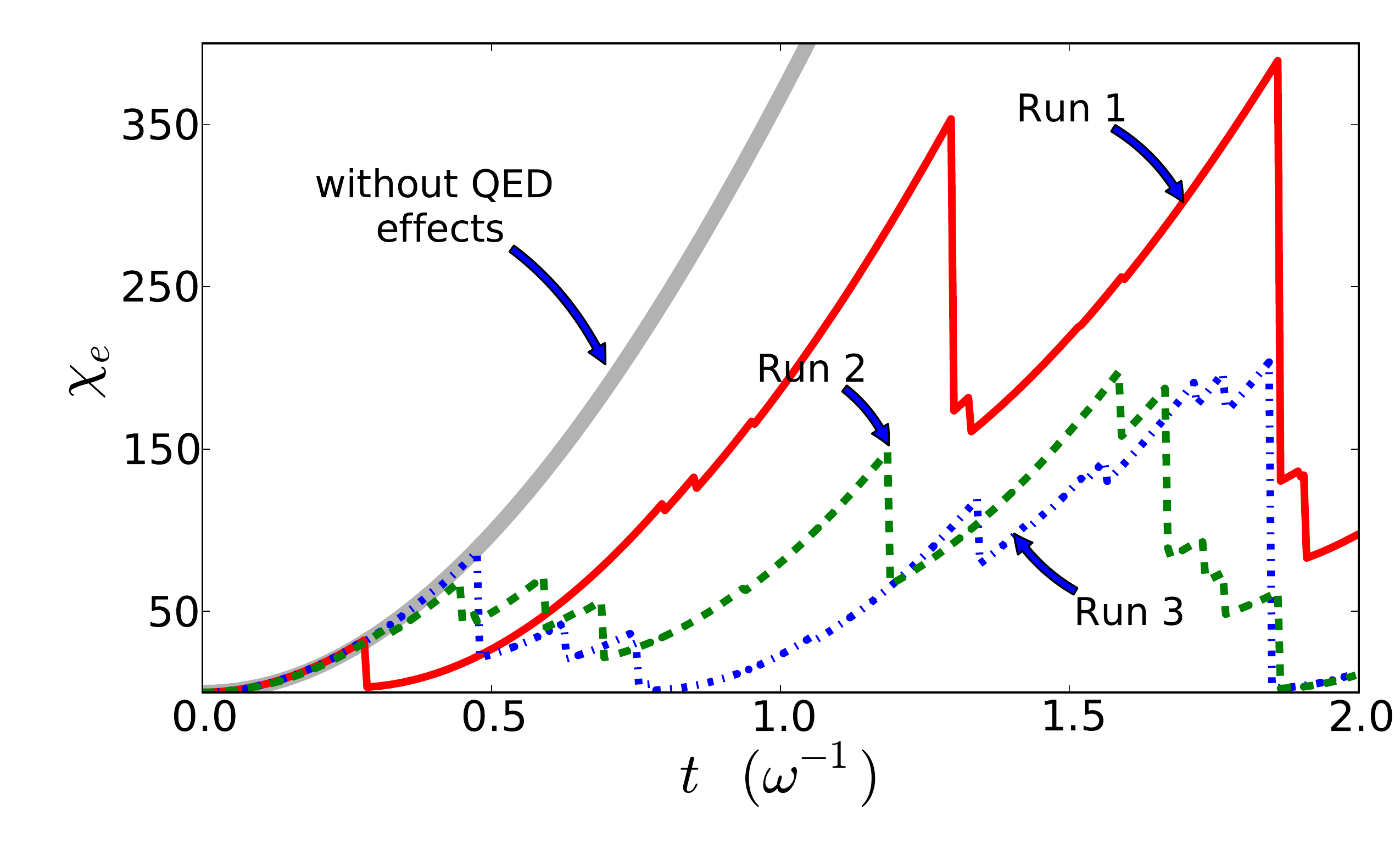}}\hspace{0.1cm}
\subfloat[]{\includegraphics[width=0.32\textwidth]{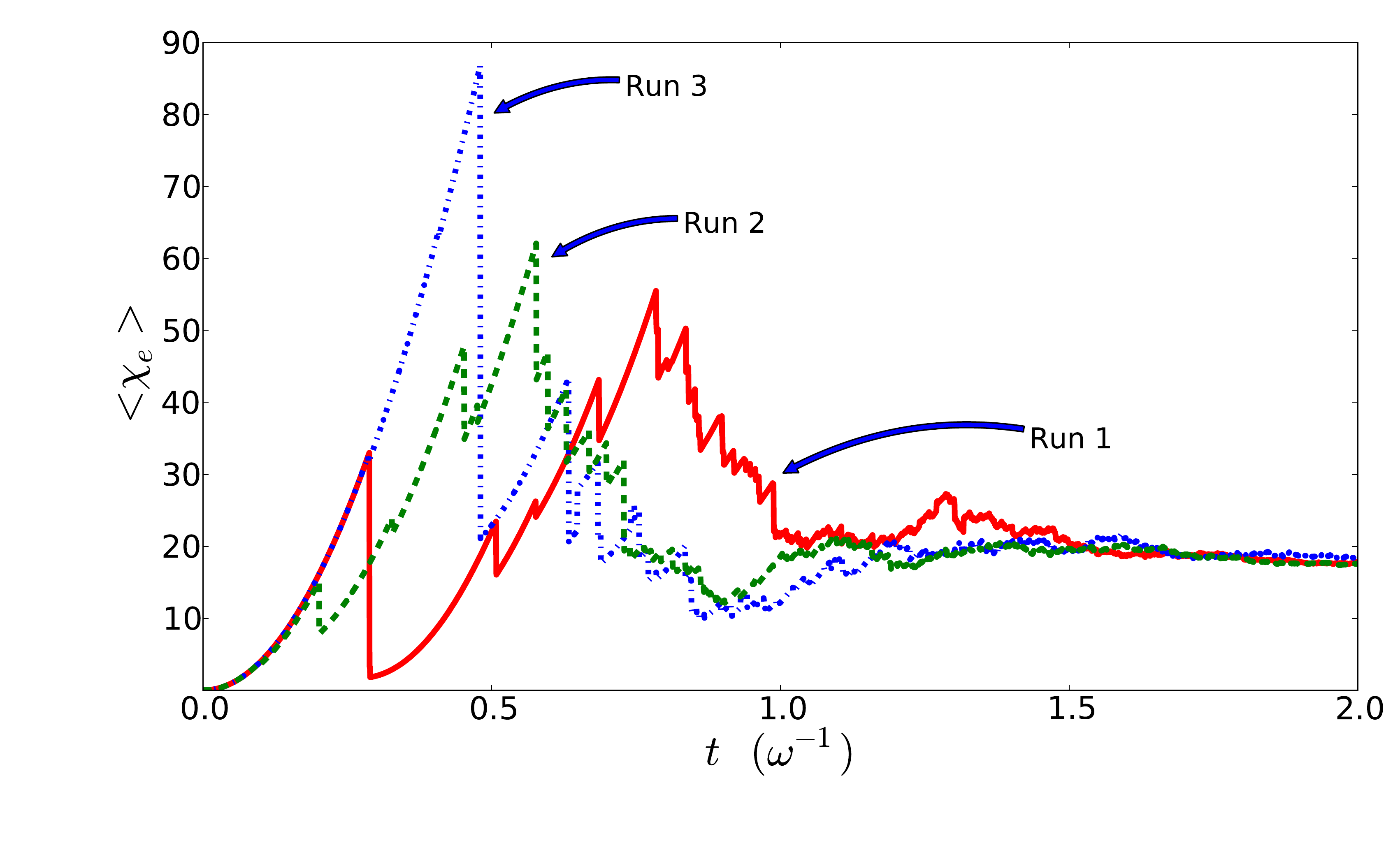}}
\caption{Simulation campaign of QED cascade generation: evolution of cascade multiplicity (a); parameter $\chi$ of the seed electron (b); and parameter $\chi$ averaged over the cascade (c) for three independent Monte Carlo runs $\at a_0=2\times 10^3$ (from \cite{Elkina2011}).}
\label{fig:cascade_sim}
\end{figure}

On the other hand, by combining Eq.~(\ref{W_q}) with Eq.~(\ref{acc_mech}) and the definition $t_{\text{e}}\simeq 1/W_\gamma(\chi(t_{\text{e}}))$ for a free path time (i.e. typical time between the QED events), in the same manner we obtain 
\begin{equation}\label{t_e}
t_{\text{e}}\simeq \frac1{\omega}\left(\frac{\alpha E_{\text{S}}}{E}\right)^{1/4},\quad \chi(t_{\text{e}})\simeq \left(\frac{E}{\alpha E_{\text{S}}}\right)^{3/2}.
\end{equation}
Obviously, for $E\gtrsim \alpha E_{\text{S}}$ (or $I\gtrsim \alpha^2 I_{\text{S}}\sim 2.5\times 10^{25}\text{W/cm}^2$) the time scales hierarchy $t_{\text{acc}}\ll t_{\text{e}}\ll 1/\omega$ is established, meaning that (i) photons are mostly emitted when $\chi\sim 1$ and hence are capable for subsequent photoproduction; and (ii) a plenty of QED events should happen on a time scale $1/\omega$ of the laser field. These are exactly the conditions singling out a QED cascade. Since due to acceleration the parameters of the particles in each generation remain the same on average, the cascade multiplicity should grow exponentially $N_{e^-e^+}\simeq e^{\Gamma t}$. This qualitative picture is fully supported by Monte Carlo simulations \cite{Legkov2010,Elkina2011,Grismayer2016}, see Fig.~\ref{fig:cascade_sim}. In particular, one can observe in Fig.~\ref{fig:cascade_sim}c how the average parameter $\chi$ tends to a definitive value, which is in fact in fairly good agreement with the estimate (\ref{t_e}). The same agreement was observed for other relevant quantities, including the increment $\Gamma\simeq 1/t_{\text{e}}$. 

The final stage of a self-sustained (A-type) cascade still remains poorly understood. For laser power $10\text{PW}$ (i.e. intensities $\sim 10^{24}\text{W/cm}^2$ and tight focusing) the duration of exponential growth of cascade multiplicity is restricted either by driving laser pulse duration or by particles escape from the focal region. However, for higher power (either higher intensity or weaker focusing) it was demonstrated that cascade multiplicity may rapidly become macroscopic \cite{Nerush2011,Grismayer2016}. When the density of created pairs exceeds the relativistic critical plasma density, the arising electron-positron plasma starts depleting the driving laser field. However, in an alternative scenario at high density the electron-positron-photon plasma may come to (quasi-)equilibrium due to various relaxation processes. Such processes, however, as of now remain almost unexplored.

\section{Radiation corrections and IFQED approach breakdown}

\begin{figure}[t]
\centering\parbox{0.8\textwidth}{\begin{multline*}
\mbox{$\frac{M}{m} =$}\underbrace{\raisebox{-0.25\height}{\includegraphics[width=1.5cm]{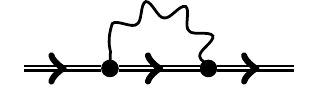}}}_{\boxed{\scriptstyle\simeq\alpha\chi^{2/3}\;\text{\cite{Ritus1972}}}}+\underbrace{\raisebox{-0.1\height}{\includegraphics[width=2cm]{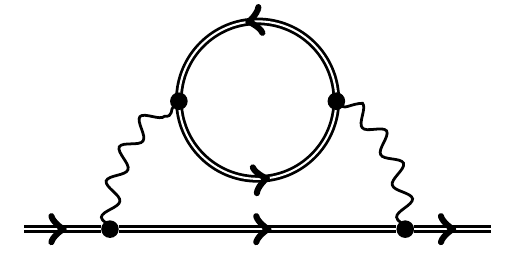}}}_{\boxed{\scriptstyle\simeq\alpha^2\chi\log{\chi}\;\text{\cite{Ritus1972}}}}+\underbrace{\includegraphics[width=2cm]{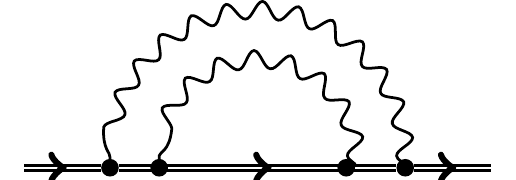}}_{\scriptstyle\simeq\alpha^2\chi^{2/3}\log{\chi}\;\text{\cite{Morozov1975}}}+\ldots+\underbrace{\includegraphics[width=2cm]{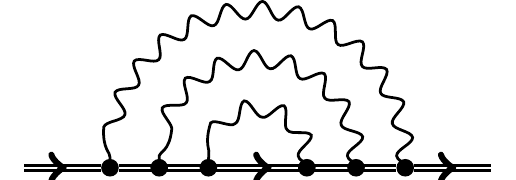}}_{\scriptstyle\simeq\alpha^3\chi^{2/3}\log^2{\chi}\;\text{\cite{Narozhny1979}}}+\\\quad\quad+\underbrace{\raisebox{-0.12\height}{\includegraphics[width=2cm]{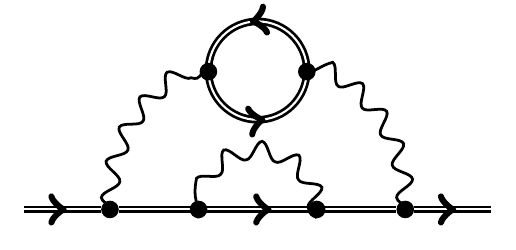}}}_{\simeq\alpha^3\chi^{4/3}\;\text{\cite{Narozhny1979}}}+\underbrace{\includegraphics[width=2cm]{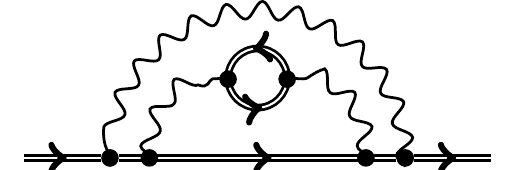}}_{\simeq\alpha^3 \chi\log^2{\chi}\;\text{\cite{Narozhny1980}}}+\underbrace{\raisebox{-0.15\height}{\includegraphics[width=2cm]{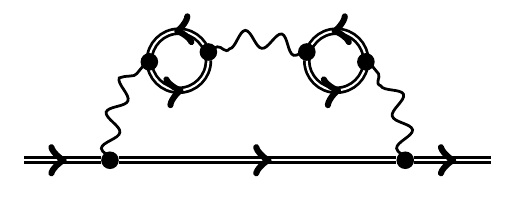}}}_{\boxed{\scriptstyle{\simeq\alpha^3\chi^{5/3}\;\text{\cite{Narozhny1980}}}}}+\ldots\\
\end{multline*}}\vspace{-1cm}
\caption{Some $2$nd and $4$th order radiation corrections to electron mass operator $M$ computed or estimated by the Ritus group in 1972-1980. The key results are enclosed in boxes.}\label{fig:R_corr}
\end{figure}

In ordinary QED, the running coupling constant $\alpha(\varepsilon\gg m)\simeq \mathcal{E}_\text{em}(\varepsilon)/m\simeq\alpha\log(\varepsilon/m)$ remains small within the whole reasonable region of energy (in particular, up to the Electroweak Theory energy scale), hence perturbation theory always works pretty well. However, as was noticed already soon after the very birth of IFQED approach\cite{Narozhny1969,Ritus1970}, the leading order contributions to the mass and polarization operators within IFQED are growing surprisingly fast with $\chi$ or $\varkappa$ (i.e. with both energy and field strength):
\begin{equation}\label{rad_corr}
M^{(2)}(\chi\gg 1)\simeq \alpha m\chi^{2/3},\quad \mathcal{P}^{(2)}(\varkappa\gg 1)\simeq \alpha m^2\varkappa^{2/3}.
\end{equation}
This can be easily traced back to Eqs.~(\ref{W_q}) via the optical theorem, and implies that for $\chi,\,\varkappa\gtrsim \alpha^{-3/2}\simeq 1.6\times 10^3$ the radiation corrections cease to be small, $M^{(2)}\simeq m$, $\mathcal{P}^{(2)}\simeq m^2$. In addition, in a proper reference frame 
\begin{equation}\label{no_rad_free}
t_{\text{e}}\sim W_{\gamma}^{-1}\simeq t_{\text{C}},\quad 
t_{\gamma}\sim W_{e^+e^-}^{-1}\simeq t_{\text{C}},
\end{equation}
meaning that the concept of radiation-free motion, obviously underlying the IFQED approach, could show up only at Compton scale, where localization is all the same impossible. If so, this should blow up the approach, thus making IFQED a truly non-perturbative theory like QCD. 

Systematic analysis of the actual expansion parameter of IFQED perturbation theory was undertaken by the Ritus group \cite{Ritus1972,Morozov1975}, and especially by Narozhny \cite{Narozhny1979,Narozhny1980}, see Fig.~\ref{fig:R_corr}. By comparing $2$-nd and $4$-th order contributions to the mass operator it was initially conjectured that the expansion parameter is $M^{(4)}/M^{(2)}\simeq\alpha\chi^{1/3}$. However, further estimation of the $6$-th order contributions identified $M^{(6)}/M^{(4)}\simeq\alpha\chi^{2/3}$ as the true expansion parameter. This conclusion is consistent with the also known data for polarization and vertex operators (not shown here for brevity). Origination of both parameters can be easily understood in terms of the qualitative approach of Sec.~\ref{sec:qual} \cite{Fedotov2016}. Indeed, the estimations (\ref{t_q}), (\ref{W_q}) were based exclusively on the uncertainty principle and thus are valid for virtual processes as well. In a proper reference frame the characteristic longitudinal and transverse sizes of a vacuum polarization\footnote{Estimations for self energy are exactly the same because ultrarelativistic kinematics is similar for both cases.} loop are estimated by $l_{\parallel,\text{P}}\simeq (m/k) t_q\simeq l_{\text{C}}\varkappa^{-2/3}$, $l_{\perp,\text{P}}\simeq \vartheta t_q\simeq eEt_q^2/k\simeq l_{\text{C}}\varkappa^{-1/3}$. Interestingly, for $\varkappa\gg 1$ both sizes turn out to be smaller than $l_{\text{C}}$ and moreover that $l_{\parallel,\text{P}}\sim r_e$ (classical electron radius!) for $\alpha\varkappa^{2/3}\sim 1$. Now the aforementioned expansion parameters can be revealed exactly as in ordinary QED as Coulomb to rest energy ratios $(e^2/l)/m$ for transverse ($l=l_{\perp,\text{P}}$) and longitudinal ($l=l_{\parallel,\text{P}}$) loop sizes, respectively. On the grounds of the optical theorem there might be also a tight relation between the higher-order radiation corrections and cascades (since cutting the former diagrams leads to the latter ones). However, for self-sustained cascades due to the scaling (\ref{t_e}) we have always $\alpha\chi^{2/3}\simeq E/E_{\text{S}}\lesssim 1$ in a view of critical field unattainability due to field depletion via cascades seeded by spontaneous pair production \cite{Fedotov2010}. Still, as pointed out above, one can attain $\alpha\chi^{2/3}\gtrsim 1$ by using an external accelerator. In fact this condition is not as exotic as one could imagine but rather is almost realizable e.g. by colliding a bunch of TeV electrons with the presently available laser pulses.

Unfortunately, in spite of obvious urgency and importance of further development of these considerations, I am not aware of any progress since the beginning of $80$'s.

\section*{Acknowledgments}
I am grateful to the organizers of the Helmholtz International Summer School ``QFT at the Limits: from Strong Fields to Heavy Quarks'' (JINR, Dubna, 2016) for inviting me to give this lecture, and to D. Blaschke, S.A. Smolyansky, A.I. Titov, A. Ilderton, T. Heinzl and A.A. Mironov for fruitful comments and inspiring discussions. The work was supported by RFBR grant 16-02-00963a.

\begin{footnotesize}

\end{footnotesize}

\begin{thebibliography}{99}
\bibitem{strickland1985compression} D. Strickland and G. Mourou, Opt. comm. {\bf 56} 219 (1985).
\bibitem{yanovsky2008ultra} V. Yanovsky \emph{et al}, Optics Express {\bf 16} 2109 (2008).
\bibitem{ELI-site}\url{www.eli-laser.eu}.
\bibitem{XCELS-site}\url{www.xcels.iapras.ru}.
\bibitem{Burke1997}D.L. Burke \emph{et al}, Phys. Rev. Lett. {\bf 79} 1626 (1997). 
\bibitem{Bamber1999}C. Bamber \emph{et al}, Phys. Rev. {\bf D60} 092004 (1999).
\bibitem{Bulanov2006}G.A. Mourou, T. Tajima, S.V. Bulanov, Rev. Mod. Phys. {\bf 78} 309 (2006).
\bibitem{diPiazza2012}A. Di Piazza, C. M\"{u}ller, K.Z. Hatsagortsyan, and C.H. Keitel, Rev. Mod. Phys. {\bf 84} 1177 (2012).
\bibitem{Narozhny2015}N.B. Narozhny, A.M. Fedotov, Contemp. Phys. {\bf 56} 249 (2015).
\bibitem{Ritus1985}V.I. Ritus, Journ. Soviet Laser Research {\bf 6} 497 (1985).
\bibitem{Fedotov2015}A.M. Fedotov, arXiv:1507.08512 (2015).
\bibitem{Mironov2014}A.A. Mironov, N.B. Narozhny, A.M. Fedotov, Phys. Lett. {\bf A378} 3254 (2014)
\bibitem{Narozhny2014}N.B. Narozhny, A.M. Fedotov, Eur. Phys. Journ. ST {\bf 223}, 1083 (2014).
\bibitem{Fedotov2010}A.M. Fedotov, N.B. Narozhny, G. Mourou, G. Korn, Phys. Rev. Lett. {\bf 105} 080402 (2010).
\bibitem{Legkov2010}M.V. Legkov, A.M. Fedotov, N. Elkina, H. Ruhl, Proc. SPIE {\bf 7994} 799423 (2010).
\bibitem{Elkina2011}N.V. Elkina \emph{et al}, Phys. Rev. STAB {\bf 14} 054401 (2011).
\bibitem{Grismayer2016}T. Grismayer \emph{et al}, Phys. Plasmas {\bf 23}, 056706 (2016).
\bibitem{Nerush2011}E.N. Nerush \emph{et al}, Phys. Rev. Lett. {\bf 106} 035001 (2011).
\bibitem{Narozhny1969}N.B. Narozhny, Sov. Phys. JETP {\bf 28} 371 (1969).
\bibitem{Ritus1970}V.I. Ritus, Sov. Phys. JETP {\bf 30} 1181 (1970). 
\bibitem{Ritus1972}V.I. Ritus, Ann. Phys. {\bf 69} 555 (1972).
\bibitem{Morozov1975}D.A. Morozov, V.I. Ritus, Nucl. Phys. {\bf B 86} 309 (1975).
\bibitem{Narozhny1979}N.B. Narozhny, Phys. Rev. {\bf D20} 1313 (1979).
\bibitem{Narozhny1980}N.B. Narozhny, Phys. Rev. {\bf D21} 1176 (1980).
\bibitem{Fedotov2016}A.M. Fedotov, arXiv:1608.02261 (2016).
\end{thebibliography}
\end{document}